\documentclass[aps, superscriptaddress,amsmath,amssymb,floatfix,bm,braket,notitlepage, twocolumn]{revtex4-1}
\usepackage{times}
\usepackage{graphicx}
\usepackage{subfigure}
\usepackage{color}
\usepackage{bm}
\usepackage{braket}
\usepackage{slashed}
\usepackage{mathtools}
\usepackage[caption=false]{subfig}
\usepackage{braket}
\usepackage{soul}

\begin{document}

\title{Glide reflection symmetry, Brillouin zone folding and superconducting pairing for the $P4/nmm$ space group}

\author{Emilian M. Nica}
\email[Corresponding author: ]{en5@rice.edu}
\affiliation{Department of Physics and Astronomy, Rice University, Houston,
Texas 77005}

\author{Rong Yu}
\affiliation{Department of Physics, Renmin University of China, Beijing 100872, China}

\author{Qimiao Si}
\affiliation{Department of Physics and Astronomy, Rice University, Houston,
Texas 77005}
\date{\today}

\begin{abstract}
Motivated by the studies of the superconducting pairing states in the iron-based superconductors,
we analyze the effects of Brillouin zone folding procedure from a space group symmetry perspective
for a general class of materials with the $P4/nmm$ space group.
The Brillouin zone folding amounts to working with an effective one-Fe unit cell, instead of the crystallographic two-Fe unit cell.
We show that the folding procedure
can be justified by the validity of a glide reflection symmetry throughout the crystallographic Brillouin zone
and by the existence of a minimal double degeneracy along the edges of the latter.
We also demonstrate how the folding procedure fails when a local spin-orbit coupling is included although the latter does not break any of the space group symmetries of the bare Hamiltonian. In light of these general symmetry considerations, we further discuss the implications of the glide reflection symmetry
for the superconducting pairing in an effective multi-orbital $t-J_{1}-J_{2}$ model. We find that the $P4/nmm$
space group  symmetry allows only pairing states with even parity  under the glide reflection and zero total momentum.
\end{abstract}

\maketitle

\section{Introduction }
\label{Sec:Space_group_no_SOC }

The iron-based superconductors form a large family of materials which exhibits considerable diversity 
in their lattice structures. Examples include the 1111, 111, and 122 iron pnictides, 
as well as the 11 iron chalcogenides. The structural unit cell of most of these superconductors consists 
of two Fe and two pnictogen/chalcogen atoms and it is typically labeled as a 2-Fe unit cell. 
In momentum space, the corresponding physical Brillouin zone (BZ) is usually referred to 
as the folded BZ (FBZ). However,
many theoretical studies have been based on tight-binding models defined on 
an effective "unfolded BZ" (UBZ) of 1-Fe unit cell
with an implicit equivalence between the former and the FBZ. The mapping between the effective 
1-Fe UBZ 
and the physical 2-Fe FBZ is via a BZ folding procedure. In view of the widespread use of this mapping 
we believe it is important to better understand the necessary conditions for its employment.

An early discussion of such a BZ folding procedure for 1111 iron-pnictide superconductors was provided 
by Lee and Wen \cite{Lee_Wen:2008}. They noticed that a single Fe-As plane contains a glide reflection symmetry 
which consists of a fractional unit cell translation followed by a reflection about the Fe-plane (see below). 
This symmetry can be used to define a pseudo-crystal momentum which can label the single-particle wave-functions 
with different parities under the operation. The last step in turn allows the definition on an unfolding procedure 
from the 2-Fe FBZ of the physical momentum space to the 1-Fe UBZ of the pseudo-crystal momentum space.  
More recently, a thorough group theoretical description of the electronic structure in the iron-based superconductors 
with a $P4/nmm$ space group symmetry has been given in Ref.~\onlinecite{Cvetkovic_Vafek:2013}. 
There, it was found that the glide-reflection symmetry classifies the Bloch states near the Fermi level and 
puts strong constraints on the low-energy effective model of the system.

The immediate motivation for the current work has come from the strong-coupling approach
to superconductivity in the multi-orbital models for the iron-based materials.
Superconducting paring in this approach has been studied by using
$t-J_1-J_2$ models with three or more $3d$ orbitals, involving at least the $3d_{xz}$, 
$3d_{yz}$ and $3d_{xy}$ set,  in an effective 1-Fe 
unit cell \cite{Goswami:2009, Yu_Nat_Comm:2013, Yu_PRB_2014};
such studies have been motivated by both the multi-orbital 
nature of the electronic structure and the bad-metal behavior of the parent compounds \cite{Yu_Nat_Comm:2013}.
As we will discuss, the Bloch states formed directly from $3d_{xz}$ and $3d_{yz}$  Wannier states 
are odd under the 
reflection about the Fe plane,
whereas those associated with the $3d_{xy}$ orbital are even 
under this symmetry operation. In the former case, the unfolding procedure becomes trivial as both types of Bloch states map onto the same quasi-crystal momentum in an unfolded BZ. By contrast, states derived directly from the $3d_{xy}$ orbital are mapped onto a quasi-crystal momentum which is shifted w.r.t. the other two orbital states. Given this, the validity of calculations done directly in a 1-Fe unit cell comes into question and a careful examination is required to establish whether a more involved 2-Fe unit cell basis should be used instead. It appears that this potential issue is
 not restricted to the particular $t-J_{1}-J_{2}$ case.
Indeed, within a more general but related context,
different ways of taking into account the two inequivalent pnictogen/chalcogen atoms have
recently been proposed ~\cite{Ong_Coleman:2013,Hu_Hao:2012}. A consensus does not appear
to be reached since even the lattice symmetries considered in these studies are quite different:
Ref.~\onlinecite{Ong_Coleman:2013} considered a $C_{4v}$ point-group  symmetry about each As atom,
while in Ref.~\onlinecite{Hu_Hao:2012} a local $S_4$ symmetry about each Fe atom was taken into account.
The proposed effective tight-binding models and the superconducting pairing symmetries
in these two works are also quite distinct. Indeed, while Ref.~\onlinecite{Ong_Coleman:2013}
proposed a spin singlet, orbital triplet $d+id$ $A_{1g}$ pairing (denoted as ``TAO pairing")
 as a consequence of the two inequivalent As atoms,
 Refs.~\onlinecite{Hu_Hao:2012,Hu:2013,Hu_Hao_Wu:2013} discussed an $s$-wave odd-parity
 pairing with nonzero total momentum,
labeled $\eta$-pairing. These proposals not only question the
 validity of using the an effective 1-Fe UBZ, they also point to the need for a clear link
 between the superconducting pairing and lattice (space group) symmetries.
 This is particularly the case for the $\eta$-pairing, where it is still uncertain whether such pairing
 is allowed by the space group symmetry ~\cite{Hu:2013,Hu_Hao_Wu:2013,Lin_Ku:2014,Wang_Maier:2014}.

We believe that part of the confusion regarding the pairing symmetry is related to the issue
we mentioned in the beginning of our paper: To what extent can a model for iron-based superconductors
defined on the 1-Fe UBZ reproduce results consistent with the one defined on a 2-Fe FBZ? Here we study
this problem for systems with a $P4/nmm$ space group (which include the 111 and 1111 iron pnictides
and the 11 iron chalcogenides). The stringent conditions imposed by the space group constitute 
our starting points in the analysis of the validity of the 1-Fe unit cell formulation. To our knowledge there have been only
a few attempts \cite{Lee_Wen:2008,Andersen_Boeri:2011,Cvetkovic_Vafek:2013,Tomic_Jeschke_Valenti:2014} 
at placing this procedure on a firmer footing.  More specifically, we wish to give a better description 
of the notion of the glide reflection symmetry within a more formal group-theoretical context and provide 
some justification for it's implicit use as a general symmetry which can be used to label Bloch states 
of arbitrary momenta.

The glide reflection operation is a part of the $P4/nmm$ space group, and is characteristic of the
non-symmorphic nature of this group. As we will see, it plays an important role in establishing,
in the absence of any spin-orbit coupling, the validity of using the 1-Fe UBZ for both the single-particle
dispersion and the pairing states. From analyzing the effect of the glide reflection, we are also able to show
that treating the local spin-orbit coupling would require working with the 2-Fe FBZ.
We should stress that we will consider the effect of the glide reflection in the context of the
entire space group symmetry.

The remainder of the paper is organized as follows. In Section \ref{Subsection:Folding_wo_SOC}, 
we aim to give a rigorous
analysis of the mapping from the physical 2-Fe BZ to an effective 1-Fe BZ for a class of materials with
$P4/nmm$ space group symmetry. Our considerations apply to the ideally 2D case for which the
conduction electrons do not disperse along the c-axis. We show that in this case, the existence of a
glide reflection symmetry for all momenta is \emph{guaranteed} by this particular space group.
Without a spin-orbit coupling, the classification of all Bloch states under the glide reflection and the
 particular degeneracies along the BZ ensure that the folding procedure does not violate any space group 
 symmetries of the system. In Section \ref{Sec:Effects_pairing} we discuss some consequences
 of the space group symmetry on the superconducting pairing in the multi-orbital $t-J_{1}-J_{2}$ model.
 In Section \ref{Sec:SOC}, we discuss the effects of an atomic spin-orbit coupling term on the
 electronic properties of both normal and superconducting states.
We first show in Section~\ref{Sec:Space_group_with_SOC} that despite the glide reflection
 still being a valid symmetry for all momenta, the lack of degeneracies along part of the BZ edge
 nullifies the usual unfolding procedure. In Sec. \ref{Sec:SOC_numerics} we present and
 compare numerical results for the normal state bandstructure with and without a spin-orbit
 coupling
 to confirm the preceding symmetry-based arguments.
 The direct consequences of the spin-orbit coupling on the pairing are then discussed 
 in Sec.~\ref{Sec:Pairing_SOC}. In Sec. \ref{Sec:Discussion} we show how the symmetry arguments 
 on the BZ folding survive for a 2D dispersion even when the interlayer couplings are taken into account. 
 We also discuss the constraint imposed by the space group symmetry on the glide-reflection parity
  of the pairing channels, and
 subsequently examine the compatibility of several proposed pairings
 with the previously described space group symmetry.
 Concluding remarks are given in Sec. \ref{Sec:Conclusion}. Appendix \ref{Sec:Appendix_A}
 contains a derivation of the important ansatz used in the main sections while related details
 are given in Appendix \ref{Sec:Appendix_B}.

\section{The $P4/nmm$ space group symmetry and the Brillouin zone folding}
\label{Sec:BZ_folding}

\begin{figure}[t!]
\includegraphics[width=1.0\columnwidth]{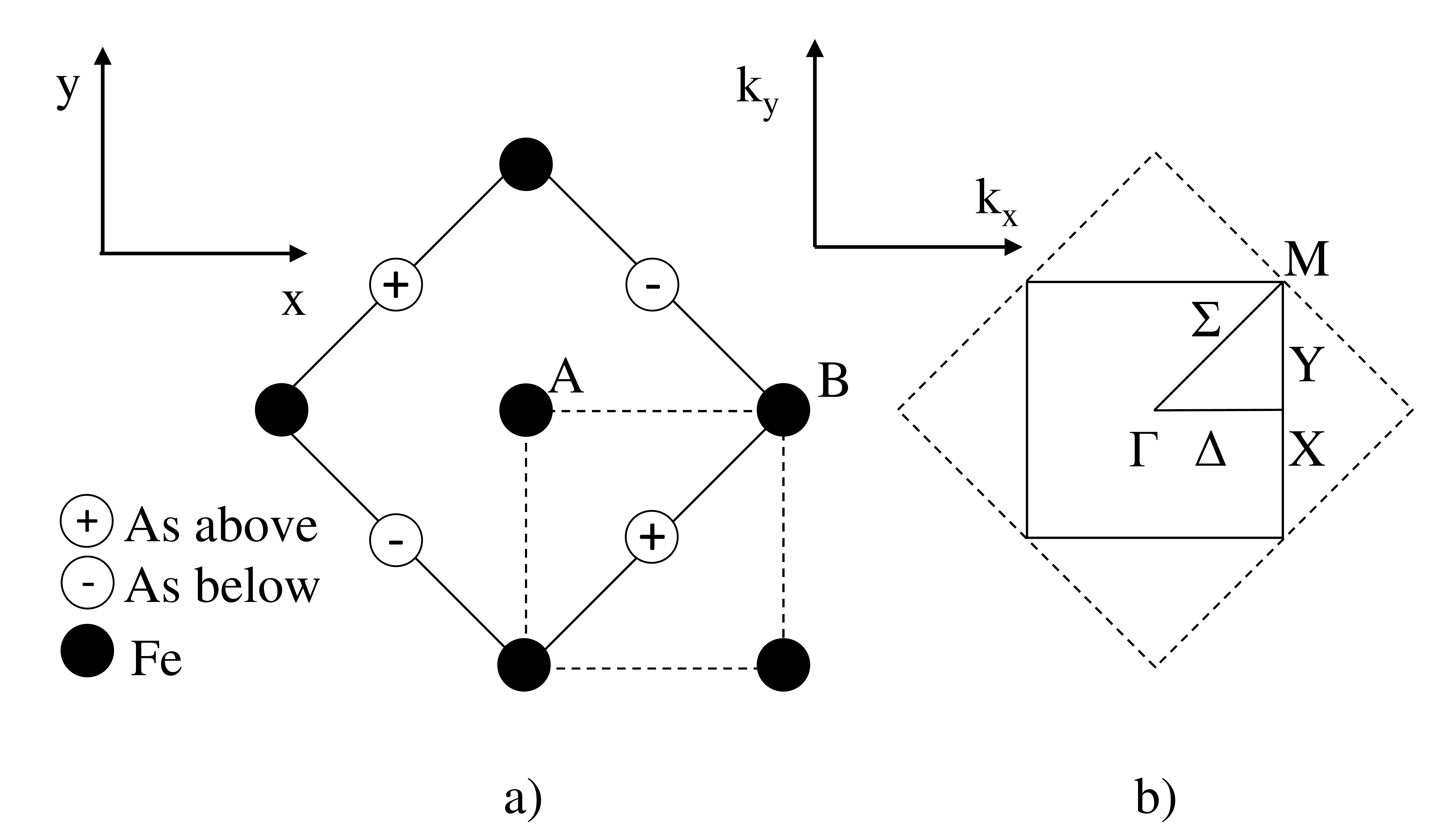}
\caption{ (a) Projection onto the Fe plane of the crystal structure of 1111 systems. The thick lines define the crystallographic 2-Fe unit cell while the dashed lines denote the effective 1-Fe unit cell. As throughout the text, the unit distance is defined by the NNN Fe translation (same sublattice). (b) The Folded Brillouin Zone (FBZ-thick line) corresponding to the true 2-Fe unit cell  and the unfolded BZ (UBZ-dashed line) of the effective 1-Fe unit cell
(for definiteness, in the $k_{z}=0$ plane). The FBZ is defined by the translation in reciprocal space by $\pmb{Q}= \left( \pm 2\pi, 0, 0 \right) $ or $\left( 0,  \pm 2\pi, 0 \right)$, which correspond to $ \left( \pm \pi, \pm \pi , 0 \right) $ or $\left( \pm \pi,  \pm \pi, 0 \right)$ in the notation for the effective 1-Fe unit cell. Here $\Gamma$, $X$ and $M$ label the points in the FBZ, while $\Delta$, $\Sigma$ and $Y$ mark the corresponding segments. }
\label{Fig:Structure_BZ}
\end{figure}

In a large group of Fe-based materials which have the $P4/nmm$ space group symmetry, the identical
Fe-pnictogen/chalcogen layers are stacked on top of each other along the c-axis.
The nontrivial spatial symmetry properties can be traced back to the structure of a single layer which is composed of a square Fe lattice in between two square As lattices shifted horizontally w.r.t. the Fe lattice and each other. The projection of the layer onto the Fe plane is shown in Fig.~\ref{Fig:Structure_BZ} (a). Two adjacent Fe sites have different nearest-neighbor As configurations defining the $A$ and $B$ sublattices.

The space group of the 1111 Fe-based superconductors is $P4/nmm$. The latter is non-symmorphic such that under any choice of unit cell one cannot decompose the set of symmetry operations into a point subgroup and its coset made up of proper lattice translations. In particular, for the conventional 2-Fe unit cell choice \cite{Int_Union:1969}, the crystal is invariant under a glide reflection symmetry $ \{ \sigma_z|\frac{1}{2} \frac{1}{2} 0 \}$ composed of a fractional unit cell translation $T_{ \pmb { \tau } } = \{ E | \frac{1}{2} \frac{1}{2} 0 \}$ in units of the sublattice translation, followed by a reflection about the Fe-plane $P_z = \{ \sigma_z| 0 0 0 \} $. The notation was chosen to be consistent with Ref.~\onlinecite{Lee_Wen:2008}.

The crucial point to consider is that the glide reflection $ T_{ \pmb { \tau } } P_z $ can be used
 to classify Bloch states of arbitrary momentum $\pmb{k}$. As observed in Ref.~\onlinecite{Cvetkovic_Vafek:2013}, for a general $ \pmb{k}=( k_{x}, k_{y}, 0) $
in the Folded Brillouin Zone (FBZ) corresponding to the 2-Fe unit cell, the group of the wave-vector
 is isomorphic to the $C_{1h} $ point group. The latter has two irreducible representations which have
 even/odd parity under the simple reflection $P_{z}$. Consequently, states belonging to the irreducible representations of the space group $P4/nmm$ for general $ \pmb{k} $ are also either Even (E) or Odd (O)
 under the glide reflection. In our view, this provides a connection between the particular glide reflection symmetry as part of the space group of the Hamiltonian and it's use in classifying states of arbitrary 2D momentum $\pmb{k}$, an argument we feel lacks an explicit exposition in the literature. By general group theoretical arguments \cite{Hammermesh:1964},
 states belonging to different irreducible representations cannot mix and thus the even/odd Bloch states
 do not hybridize. As we show in the following, this, together with the particular degeneracies along the
 entire folded BZ edge
 (see Fig. \ref{Fig:Structure_BZ}) allows the reduction to the 1-Fe unit cell.

\noindent 

\subsection{The glide reflection symmetry and the Brillouin Zone folding}
\label{Subsection:Folding_wo_SOC}

The advantage of using an unfolded 1-Fe BZ consists in effectively reducing the number of bands in the calculation. In practice this means introducing a set of trial wavefunctions defined on an 
UBZ, carrying on the calculation, then folding back to the physical FBZ. In doing this, one must check that the Bloch functions defined in the UBZ do not introduce additional hybridization terms in the Hamiltonian, and that they do not violate any of the original symmetries of the space group.  Here, we analyze the same procedure in reverse. 
We start with a set of trial wave functions which are in accord with a minimum space group symmetry, 
namely the glide reflection, then examine the unfolding process.

As mentioned in the previous section, the glide-reflection can be used to classify the irreducible representations of the space group for general $\pmb{k}$ in the folded BZ when there is no dispersion along the $z$-axis i.e. $k_{z}=0$. This ensures that eigenstates of this operator will be eigenstates of the Hamiltonian within a transformation on the orbital indices alone.  As a consequence, the Hamiltonian can be written in block diagonal form at these general $\pmb{k}$. 
The detailed representation theory of this space group shows that for a number of higher-symmetry momenta 
in the FBZ, the irreducible representations are not required to be eigenstates of the glide \cite{Cvetkovic_Vafek:2013}. This does not invalidate our arguments since by continuity any off-diagonal terms must vanish here as well.  
We will show that this implies a one-to-one correspondence between the eigenstates of the glide-reflection and 
Bloch states defined on the UBZ, with no hybridization between different momenta.

We proceed to build Bloch eigenstates of the glide operation with $\pmb{k}$ in the folded zone (See Appendix \ref{Sec:Appendix_A}). The electron annihilation operator in the physical 2-Fe BZ on sublattice A(B) is defined as

\noindent \begin{align}
\label{Eq:sublattice}
C_{\pmb{k},A/B, \alpha}=\frac{1}{\sqrt{N_{s}}} \sum_{i} e^{i \pmb{k} \cdot \pmb{R}_{i,A/B}} C_{\pmb{R}_{i,A/B}, \alpha},
\end{align}

\noindent where $\alpha$ is an orbital index, $N_s=N/2$ refers to the number of (2-Fe) unit cells and $i$ is the index of the unit cell. The two sublattice vectors are related by $\pmb{R}_{i,A} + \pmb{\tau} = \pmb{R}_{i,B}$ with $\tau$ the nearest-neighbour distance used in the translation of the glide operation.  The anihilation operators do not have definite parity under the glide reflection $T_{ \pmb { \tau } } P_z$:

\noindent \begin{align}
(T_{ \pmb { \tau } } P_z) C_{\pmb{k},A/B, \alpha} = (-1)^{\alpha} 
C_{\pmb{k},B/A, \alpha},
\end{align}

\noindent where 
$(-1)^{\alpha} =\pm 1$
depending on the parity of the local orbital under a pure reflection. 
Without loss of generality we can define operators with definite parity :

\noindent \begin{align}
C_{\pmb{k},E,\alpha} = & \frac{1}{\sqrt{2}} \left[ C_{\pmb{k},A,\alpha} + (-1)^{\alpha}  C_{\pmb{k},B,\alpha} \right] \label{Eq:Eigen_1_a} \\
C_{\pmb{k},O,\alpha} = & \frac{1}{\sqrt{2}} \left[ C_{\pmb{k},A,\alpha} - (-1)^{\alpha}  C_{\pmb{k},B,\alpha} \right]. \label{Eq:Eigen_1_b}
\end{align}

\noindent
Note
that the even ($E$) and odd ($O$) parity states refer to parity under the glide and so we can build both types of operators for arbitrary orbital parity $\alpha$. We also refer the reader to Appendix \ref{Sec:Appendix_A} for more details on the states defined above.

Since $C_{\pmb{k},E,\alpha}$ and $C_{\pmb{k},O,\alpha}$  have different parity under the glide for arbitrary "2-Fe" crystal momentum , they do not mix in a one-particle Hamiltonian. An arbitrary 2D tight-binding Hamiltonian consistent with the space group symmetries can then be expressed in the terms of the $E/O$ states as

\noindent \begin{align}
\label{Eq:2_Fe_Hamiltonian}
H_{TB} = \sum_{\pmb{k} \in FBZ } \left( \epsilon_{E}^{\alpha \beta} (\pmb{k}) C_{\pmb{k},E,\alpha}^{\dagger} C_{\pmb{k},E,\beta} + \epsilon_{O}^{\alpha \beta} (\pmb{k}) C_{\pmb{k},O,\alpha}^{\dagger} C_{\pmb{k},O,\beta}\right),
\end{align}

\noindent where $FBZ$ stands for the 2-Fe Folded Brillouin Zone and $\epsilon_{E}^{\alpha \beta}(\pmb{k}), \epsilon_{O}^{\alpha \beta} (\pmb{k})$ are matrices in orbital space. We omitted the spin index for simplicity.

We now turn to the unfolding procedure and start by noting that the eigenstates of the glide reflection have (see Eqs. \ref{Eq:App_state_a} and \ref{Eq:App_state_b})

\noindent \begin{align}
\label{Eq:Irr_switch}
C_{\pmb{k}+\pmb{Q},E/O , \alpha} = C_{\pmb{k},O/E, \alpha},
\end{align}

\noindent
where $\pmb{Q}=(\pm 2 \pi, 0)$ or $\pmb{Q}=(0, \pm 2\pi)$
 in units of the 2-Fe unit cell, and correspond to
 $ \left( \pm \pi, \pm \pi , 0 \right) $ or $\left( \pm \pi,  \pm \pi, 0 \right)$ in units of the
 effective 1-Fe unit cell. That is, our states apparently violate the symmetry under the pure translation by a 2-Fe unit cell sublattice. In the simplest Bravais lattice case a state of arbitrary momentum is only labeled by the irreducible representation of the pure translation subgroup i.e by $\pmb{k}$ itself. Here, in addition to $\pmb{k}$ one can also label all the states according to the irreducible representations of the glide reflection as well, that is, by both $\pmb{k}$ and $\lambda$. The resolution to our apparent conundrum lies in the fact that states displaced by a reciprocal lattice vector $\pmb{Q}$ can switch their representation with respect to the glide reflections:
\noindent \begin{align}
\label{Eq:E_to_O}
E \rightarrow O, \epsilon_{E}^{\alpha \beta} (\pmb{k}+ \pmb{Q}) = \epsilon_{O}^{\alpha \beta} (\pmb{k}).
\end{align}
This ensures that we are not violating the original pure translation symmetry but it also introduces a constraint: Eigenstates of the system must be at least doubly degenerate all along the edge of the first
 (folded) BZ. Only in such case can the states switch representations as one goes beyond the folded BZ.  Indeed, this is precisely what happens in Si and Ge at the X point along the edge of the BZ \cite{Slater_1954}, \cite{Dresselhaus_1968}.  The lattice in this case has a diamond structure and the two sublattices can be connected by a glide reflection. Due to the non-symmorphic nature of the space group, the states are degenerate at X and must exchange representations as one goes into the extended BZ. In the present case for $P4/nmm$, there is a minimal degeneracy throughout the edge of the folded BZ \cite{Cvetkovic_Vafek:2013}, \cite{Cracknell_book:1972} which allows for the switching of the representations as one crosses into the extended zone.

We can connect with one of the initial folding procedures \cite{Lee_Wen:2008} which relied on a definition of a quasi-crystal-momentum dependent on the parity of the orbital state under the pure reflection. Using the equivalent form of the glide states in (\ref{Eq:App_state_a}), (\ref{Eq:App_state_b}), we can absorb the negative sign from $(-1)^{(\alpha=1)}=e^{\pm i \pmb{Q \cdot \tau }}=-1$ for states with odd parity under a pure reflection. Here $\pmb{Q}=(\pm 2 \pi, 0)$ or $\pmb{Q}=(0, \pm 2\pi)$ and $\tau=( \pm \frac{1}{2},\pm \frac{1}{2} )$ in units of the 2-Fe unit cell. Explicitly, the eigenstates of the glide become

\noindent \begin{align}
C_{\pmb{k},E,\alpha ~\text{even}}& = & C_{\pmb{k}, \alpha} & = & \frac{1}{\sqrt{N}}  \sum_{\pmb{R}_i} \left[ e^{i \pmb{k} \cdot \pmb{R}_{i} } C_{\pmb{R}_i, \alpha } \right] \label{Eq:Orb_Fourier_a} \\
C_{\pmb{k},E,\alpha~\text{odd}}& = &  C_{\pmb{k}+ \pmb{Q}, \alpha} & =  & \frac{1}{\sqrt{N}}  \sum_{\pmb{R}_i} \left[ e^{i ( \pmb{k} + \pmb{Q}) \cdot \pmb{R}_{i} } C_{\pmb{R}_i, \alpha} \right] \label{Eq:Orb_Fourier_b} \\
C_{\pmb{k},O,\alpha~\text{even}}& = & C_{\pmb{k}+ \pmb{Q}, \alpha} & = & \frac{1}{\sqrt{N}}  \sum_{\pmb{R}_i} \left[ e^{i (\pmb{k} + \pmb{Q} ) \cdot \pmb{R}_{i} } C_{\pmb{R}_i, \alpha} \right] \label{Eq:Orb_Fourier_c} \\
C_{\pmb{k},O,\alpha~\text{odd}}& = & C_{\pmb{k}, \alpha} & = & \frac{1}{\sqrt{N}}  \sum_{\pmb{R}_i} \left[ e^{i \pmb{k} \cdot \pmb{R}_{i} } C_{\pmb{R}_i, \alpha} \right] \label{Eq:Orb_Fourier_d}.
\end{align}

\noindent  The expressions above can be formally subtituted into the Hamiltonian (\ref{Eq:2_Fe_Hamiltonian}). Under this mapping, the initial summation over $\pmb{k} \in $ FBZ for both $ E $ and $ O $ glide parity sectors can be re-written as a summation over $\slashed{\pmb{k}} \in $ UBZ (unfolded BZ) for the $E$ only sector (recall that $ \epsilon_{\alpha \beta}^{E} (\pmb{k}+ \pmb{Q}) = \epsilon_{\alpha \beta}^{O} (\pmb{k}) $). By taking into account the fact that different \emph{orbital} parity states must be displaced by $\pmb{Q}$ w.r.t. to each other we can recast the tight-binding Hamiltonian as

\noindent \begin{widetext}
\begin{align}
H_{TB} = & \sum_{\pmb{\slashed{k}} \in UBZ} \Bigg[ \sum_{ee} \left( \epsilon^{E}_{\alpha \beta}(\pmb{\slashed{k}}) C^{\dagger}_{\pmb{\slashed{k}}, \alpha} C_{\pmb{\slashed{k}}, \beta} \right) + \sum_{oo} \left( \epsilon^{E}_{\alpha \beta}(\pmb{\slashed{k}}) C^{\dagger}_{\pmb{\slashed{k}}+\pmb{Q}, \alpha} C_{\pmb{\slashed{k}}+\pmb{Q}, \beta} \right)
  + \sum_{eo} \left( \epsilon^{E}_{\alpha \beta}(\pmb{\slashed{k}}) C^{\dagger}_{\pmb{\slashed{k}}, \alpha} C_{\pmb{\slashed{k}}+\pmb{Q}, \beta}  \right) + \sum_{oe} \left( \epsilon^{E}_{\alpha \beta}(\pmb{\slashed{k}}) C^{\dagger}_{\pmb{\slashed{k}}+\pmb{Q}, \alpha} C_{\pmb{\slashed{k}}, \beta} \right) \Bigg]  \label{Eq:Conventioanal_1}\\
= &  \sum_{\pmb{\tilde{k}} \in UBZ}  \sum_{\alpha \beta} \left( \epsilon^{E}_{\alpha \beta}(\pmb{\tilde{k}}) C^{\dagger}_{\pmb{\tilde{k}}, \alpha} C_{\pmb{\tilde{k}}, \beta} \right),
\end{align}
\end{widetext}

\noindent where $e,o$ refer to the parity of the orbitals $\alpha$ and $\beta$. The details of the above derivation are presented in Appendix \ref{Sec:Appendix_B}. In going to the second line we explicitly identified along with Ref. \onlinecite{Lee_Wen:2008} the dependence of $\slashed{\pmb{k}}$ on the orbital parity such that $\pmb{\tilde{k}}=\pmb{\slashed{k}}$ for even orbital states and $\pmb{\tilde{k}}= \pmb{\slashed{k}}+ \pmb{Q}$ for odd states. What allows us to equate the two forms is the fact that $\pmb{\slashed{k}}$ is effectively a dummy variable in (\ref{Eq:Conventioanal_1}) and the matrix structure is determined by the orbital parities alone.
For the purpose of notational clarity, we stress the following: a) $\pmb{\slashed{k}}$ is defined in the larger 
Brillouin zone (UBZ) for the 1-Fe unit cell, and is associated with an orbital-dependent procedure 
in unfolding from the smaller Brillouin zone (FBZ). It has the same meaning as that used in 
Ref.~\onlinecite{Lee_Wen:2008}; b) $\pmb{\tilde{k}}$, also defined in the UBZ for the 1-Fe unit cell, 
arises in an orbital-independent unfolding procedure. The distinction between these two wave vectors 
will be important for our later considerations.

We can illustrate a different widely used unfolding procedure if we define two types of local states. 
Specifically, we can set $C_{\pmb{R_{A}}, \alpha}= C_{\pmb{R_{A}}+ \pmb{\tau}, \alpha}$ 
for even parity orbital
 $\alpha$ and $C_{\pmb{R_{A}}, \alpha}= (-1)C_{\pmb{R_{A}}+ \pmb{\tau}, \alpha}$ for odd 
 parity orbital $\alpha$ in Eqs. \ref{Eq:Orb_Fourier_a}-\ref{Eq:Orb_Fourier_d}  
 such that the local orbital state changes sign under the simple 1-Fe unit cell translation $T_{\tau}$. 
 The additional minus sign for odd orbital states will generate and additional $e^{i\pmb{Q}\cdot \pmb{\tau}}$ 
 factor for odd orbital states. This is equivalent to the orbital parity dependent gauge transformation encountered
  in the literature. Using the same notation as above the eigenstates become

\noindent \begin{align}
C_{\pmb{k},E,\alpha} =& C_{\pmb{k}, \alpha} \label{Eq:Gauge_a} \\
C_{\pmb{k},O, \alpha} = & C_{\pmb{k}+\pmb{Q}, \alpha} \label{Eq:Gauge_b}
\end{align}

\noindent for arbitrary parity $\alpha$.  If we identify $\epsilon^{O}_{\alpha \beta}(\pmb{k})= \epsilon^{E}_{\alpha \beta}(\pmb{k}+\pmb{Q})$ and use the above definitions in the Hamiltonian (\ref{Eq:2_Fe_Hamiltonian}) the latter can be re-written as

\noindent \begin{align}
H_{TB} =&  \sum_{\pmb{k} \in FBZ } \bigg( \epsilon_{E}^{\alpha \beta} (\pmb{k}) C_{\pmb{k},\alpha}^{\dagger} C_{\pmb{k},\beta} \notag \\
& + \epsilon_{E}^{\alpha \beta} (\pmb{k}+\pmb{Q}) C_{\pmb{k}+\pmb{Q},\alpha}^{\dagger} C_{\pmb{k}+\pmb{Q},\beta}\bigg) \\
= & \sum_{\pmb{\tilde{k}} \in UBZ }\epsilon_{E}^{\alpha \beta} (\pmb{\tilde{k}})C_{\pmb{\tilde{k}},\alpha}^{\dagger} C_{\pmb{\tilde{k}},\beta}
\end{align}

\noindent where $\pmb{\tilde{k}}$ is in the Unfolded Brillouin Zone (UBZ).

We stress that both unfolding procedures discussed above were guaranteed by the absence of mixing 
between the $E,O$ sectors in the Hamiltonian for general $\pmb{k}$ together with the switching of the glide 
parity for the ansatz states beyond the first folded BZ (Eq. \ref{Eq:Irr_switch}). These two conditions are a direct consequence of the irreducible representations of the glide symmetry for arbitrary $\pmb{k}\in $ FBZ. 
None of the folding procedures discussed above and used in the literature would work if the Hamiltonian 
contained $E,O$ mixing terms since we could not define $\pmb{Q}$ shifted states related to the 1-Fe unit cell 
Fourier transforms for either even and odd parity orbitals.

Although the above arguments seem to be implicit in the literature \cite{Andersen_Boeri:2011, Lee_Wen:2008},
we have encountered  few observations of the crucial connection to the irreducible representations
of the $P4/nmm$ space group, with a notable exception of Ref. \onlinecite{Tomic_Jeschke_Valenti:2014} in the context of the electronic structure.
Here, by constructing the electronic states which are even or odd under glide reflection, we are in position to discuss
the  the constraints imposed by the glide-reflection symmetry on the various types of pairing, which will be discussed
 in the next subsection, as well as the effects of a local spin-orbit coupling on the unfolding procedure,
 which we turn to in the following section. As we mentioned in Sec. \ref{Sec:BZ_folding},
 we believe that it is insufficient to argue that the presence of a glide reflection operation in the space
 group of the Hamiltonian guarantees the success of a folding procedure. Along with the glide reflection,
 the space group contains fifteen other symmetry elements \cite{Int_Union:1969} different from pure sublattice translations. Of these, only the glide reflection can be used to classify states according to the irreducible
 representations of the space group for \emph{general two-dimensional crystal momenta},
 since only this operation leaves an arbitrary $\pmb{ k } $ invariant. But as we will show in Sec.~\ref{Sec:SOC},
 the same folding procedure fails when a spin-orbit coupling is turned on, even though the additional interactions
 do not violate the glide reflection symmetry or any space symmetry of the original Hamiltonian.

\subsection{The effects of the glide symmetry on the superconducting pairing interactions}
\label{Sec:Effects_pairing}

We now discuss some of the effects of the glide symmetry for the superconducting pairing interaction. 
We assume that the symmetry outlined in the previous sections is not violated such that without the pairing 
interactions one can unfold the BZ of the 2-Fe unit cell to the effective 1-Fe UBZ. We also assume 
that the original symmetry of the 2-Fe unit cell is not broken by the appearance of magnetic or ``nematic" order. 
The analysis presented here also assumes a trivial spatial dependence of the bare pairing interactions such as 
would arise in a nearest neighbor (NN), next-nearest neighbor (NNN) $t-J_{1}-J_{2}$ Hamiltonian. 
Lastly, we consider singlet pairing, as evidenced by experiments for various iron-based superconductors. In this case, the antisymmetric nature of the pairing wavefunction requires the pairing 
to be even parity under inversion~\cite{Sigrist_Ueda:1991}.

Before proceeding in a manner analogous to that of Sec. \ref{Subsection:Folding_wo_SOC}, 
some additional remarks are 
in order. The tight-binding part of a Hamiltonian with pairing interactions at mean-field level is typically 
chosen as the identity representation of the space group. In general, the pairing functions are determined 
self-consistently and can lower the symmetry of the Hamiltonian to a subgroup of the full space group (as {\it e.g.}, what happens to the rotational invariance of the Hamiltonian under the $C_4$ operation of the $D_{4h}$ point group of the full $P4/nmm$ space group in the case of a d-wave pairing).
The problem simplifies for the spin singlet pairing considered in this section. For Cooper pairs of equal and opposite momenta in the FBZ ( $\pmb{k}, -\pmb{k}$) 
associated with the 2-Fe unit cell,
the symmetry properties 
of the pairing functions are completely determined by the $D_{4h}$ point group associated 
with the $P4/nmm$ space group. More precisely, the tensor irreducible  representations 
of the space group used to classify the pairing are the irreducible representations of the group 
of the wave-vector labeled by total momentum $\pmb{k}+(-\pmb{k})=(0,0)$  
(For a general argument, see Ref.~\onlinecite{Ozaki:1995}).
Given this, we know that the irreducible representations of the $D_{4h}$ point group are either even 
or odd under inversion. In addition, for zero total momentum representations, the glide reflection
 is equivalent to a simple reflection about the z-plane. This means that the parity under inversion coincides 
 with the parity under the glide-reflection for one dimensional irreducible tensor representations and is opposite 
 for the two-dimensional ones~\cite{Cvetkovic_Vafek:2013}.  Restricting ourselves to the former 1D  
representations, a Hamiltonian containing any linear combination of inversion-even pairing terms cannot 
break the glide reflection symmetry,
regardless of whether the pairing preserves the rotational invariance or not.
The glide symmetry can
be broken if and only if the system spontaneously breaks the
 inversion symmetry, which consequently results in triplet pairing. 
 The remaining part of the section seeks to illustrate that no finite momentum pairing terms 
 can be present in either the folded or unfolded BZ provided that the Hamiltonian 
is invariant under the glide-reflection. 
 This also guarantees that the common folding 
 procedure is still valid in this case.

We consider for illustration purposes the following NN pairing interaction:

\noindent \begin{align}
\label{Eq:Pairing_interaction}
H_{int}=\sum_{\alpha \beta}\sum_{ij,NN} J_{\alpha \beta} & \left[ C^{\dagger}_{i \alpha \uparrow}C^{\dagger}_{j \beta \downarrow} - C^{\dagger}_{i \alpha \downarrow}C^{\dagger}_{j \beta \uparrow} \right] \times \notag \\
& \left[ C_{i \alpha \uparrow}^{\phantom{\dagger}} C_{j \beta \downarrow} - C_{i \alpha \downarrow}C_{j \beta \uparrow} \right]+H.C.
\end{align}

Based on Eqs. \ref{Eq:Eigen_1_a}, \ref{Eq:Eigen_1_b} we can define Fourier transforms on each sublattice as

\noindent \begin{align}
C_{\pmb{R}_{A}, \alpha }  &  = \frac{1}{ \sqrt{ N_s } } \sum_{\pmb{k } \in FBZ} e^{-i \pmb{k} \cdot \pmb{ R_{A} } } \left[ C_{\pmb{ k },E,\alpha} + C_{\pmb{ k },O,\alpha} \right], \label{FT_A}  \\
C_{\pmb{R}_{B}, \alpha}  &  = (-1)^{\alpha} \frac{1}{ \sqrt{ N_s } } \sum_{\pmb{k } \in FBZ} e^{-i \pmb{k} \cdot \pmb{ R_{B} } } \left[ C_{\pmb{ k },E,\alpha} - C_{\pmb{ k },O,\alpha} 
\right]. \label{FT_B}
\end{align}

\noindent It is straightforward to see that the pairing interaction in Eq.\ref{Eq:Pairing_interaction} will contain equal numbers of $E$ and $O$ states. Indeed, this can be understood from the general discussion in the previous sections. Although formally at some special points of the FBZ the eigenstates are not required to also be eigenstates of the glide reflection, in practice, continuity must enforce this as was mentioned in the preceding section. All terms in the Hamiltonian must effectively be invariant under the glide reflection symmetry for arbitrary $\pmb{k}$, ensuring that all interactions must be even under the glide reflection symmetry. This constrains all pairing terms to have equal numbers of even and odd states by the space group symmetry arguments. The general form of the interactions can in principle generate pairing terms mixing $E$ and $O$ states which might result in a finite $\pmb{Q}$ momentum Cooper pairs.

To make progress, we can consider (\ref{Eq:Pairing_interaction}) in a mean-field (MF) approach. In view of the connection between the MF Hamiltonian and the equation of motion approach we anticipate the same results beyond the simplest level. As an illustration we analyze the term

\noindent \begin{align}
\label{Eq:MF_term}
& H'_{int, MF} = \sum_{\alpha\beta} \sum_{<ij>}\sum_{ e } \notag \\
& \left< J_{\alpha \beta} \left[ C^{\dagger}_{i \alpha \uparrow}C^{\dagger}_{j \beta \downarrow} - C^{\dagger}_{i \alpha \downarrow}C^{\dagger}_{j \beta \uparrow} \right] \right > \times C_{i \alpha \uparrow}C_{j \beta \downarrow}.
\end{align}

\noindent The remaining terms in Eq.\ref{Eq:Pairing_interaction} can be obtained by flipping the spin indices of the last pair and adding the Hermitian conjugate terms. Decomposing the NN summation as $\sum_{<ij>}\equiv \sum_{\pmb{R_{A}}}\sum_{\pmb{e}=\pm \pmb{\hat{x}, \hat{y}}}+\sum_{\pmb{R_{B}}}\sum_{\pmb{e}=\pm \pmb{\hat{x}, \hat{y}}}$ and taking the sublattice specific 
Fourier transformation (F.T.) as in (\ref{FT_A}),(\ref{FT_B}) we obtain

\noindent \begin{widetext}
\begin{align}
\label{Eq:Pairing_MF}
H'_{int, MF} =&  \sum_{\alpha\beta} \sum_{ \pmb{e} } \sum_{\pmb{k} \in FBZ} e^{ i\pmb{ke} } \times \notag \\
 \Bigg[ & (-1)^{\beta} \Delta^{A}_{\pmb{e},\alpha\beta} ( C_{\pmb{k},E, \alpha \uparrow}C_{\pmb{-k},E, \beta \downarrow} - C_{\pmb{k},E, \alpha \uparrow}C_{\pmb{-k},O, \beta \downarrow} +  C_{\pmb{k},O, \alpha \uparrow }C_{\pmb{-k},E, \beta \downarrow} - C_{ \pmb{k},O, \alpha \uparrow}C_{ \pmb{-k},O,\beta \downarrow} )  + \notag \\
+ &  (-1)^{\alpha}  \Delta_{\pmb{e},B,\alpha\beta}^{B} (C_{\pmb{k},E, \alpha \uparrow}C_{\pmb{-k},E, \beta \downarrow} + C_{\pmb{k},E, \alpha \uparrow }C_{-k \beta \downarrow O} -  C_{\pmb{ k},O, \alpha \uparrow}C_{ \pmb{-k},E, \beta \downarrow } - C_{ \pmb{k},O, \alpha \uparrow}C_{\pmb{-k},O, \beta \downarrow } ) \Bigg]
\end{align}
\end{widetext}

\noindent where

\noindent \begin{align}
\label{Eq:Delta_A}
\Delta^{A}_{\pmb{e},\alpha\beta}=&  \left< J_{\alpha \beta} \left[ C^{\dagger}_{\pmb{ R_{A}} \alpha \uparrow}C^{\dagger}_{\pmb{ R_{A}}+\pmb{e} \beta \downarrow} - C^{\dagger}_{\pmb{ R_{A}} \alpha \downarrow}C^{\dagger}_{\pmb{ R_{A}}+\pmb{e} \beta \uparrow} \right] \right >
\end{align}

\noindent \begin{align}
\label{Eq:Delta_B}
\Delta^{B}_{\pmb{e},\alpha\beta}=&  \left< J_{\alpha \beta} \left[ C^{\dagger}_{\pmb{ R_{B}} \alpha \uparrow}C^{\dagger}_{\pmb{ R_{B}}+\pmb{e} \beta \downarrow} - C^{\dagger}_{\pmb{ R_{B}} \alpha \downarrow}C^{\dagger}_{\pmb{ R_{B}}+\pmb{e} \beta \uparrow} \right] \right >
\end{align}

The crucial assumption made in the above equations is that both $\Delta^{A}_{e}$, $\Delta^{B}_{e}$ are independent of their sublattice space indices $\pmb{R_{A}}$ and $\pmb{R_{B}}$, but $\Delta^{A}_{e} \neq \Delta^{B}_{e}$ in general. This amounts to having the pairing order parameters which are constant on the respective sublattices.

We distinguish two cases: i) $\alpha$ and $\beta$ corresponding to orbitals of the same parity under the reflection, ii) $\alpha$ and $\beta$ have different parities.

For case i) of same parity we can look at the real space expression (\ref{Eq:MF_term}) and demand invariance under a glide reflection. Since both $\alpha$, $\beta$ terms acquire the same orbital parity factor and each state gets shifted by one unit of the 1-Fe unit cell we have $\Delta^{A}_{\pmb{e},\alpha\beta}=\Delta^{B}_{\pmb{e},\alpha\beta}$. Plugging this into (\ref{Eq:Pairing_MF}) and setting $(-1)^{\alpha}=(-1)^{\beta}$, we see that all $EO$ and $OE$ mixed terms cancel and we get only same glide parity terms.

For case ii) where $\alpha$ and $\beta$ have different orbital parities we can do the same as above and impose the glide reflection symmetry in real space. We get  $\Delta^{A}_{\pmb{e},\alpha\beta}=-\Delta^{B}_{\pmb{e},\alpha\beta}$. With $(-1)^{\alpha}=-(-1)^{\beta}$ we get the same cancellation as for case i).

Identical results are obtained for the next-nearest coupling (NNN, $J_2$) case.

In both cases of same and different orbital parity pairing, the results are completely analogous to those considered in Section \ref{Subsection:Folding_wo_SOC} for a tight-binding model without spin-orbit coupling. Namely, the pairing part of the Hamiltonian splits into $EE$ and $OO$ glide parity sectors. The exact same unfolding arguments can be trivially extended to the pairing part. Calculations can be done in an UBZ using the $\pmb{\tilde{k}}$  F.T. and then fold the results to the 2-Fe BZ. In this scheme there is no finite momentum pairing.

\section{The effects of spin-orbit coupling}\label{Sec:SOC}
\subsection{Folding in the presence of spin-orbit coupling}
\label{Sec:Space_group_with_SOC}

A local (atomic) spin-orbit coupling (SOC) term
\begin{equation}
\label{Eq:SOC_term}
I\pmb{L} \cdot \pmb{S} = \frac{I}{2} \left ( \pmb{J}^{2} - \pmb{L}^{2} -  \pmb{S}^{2} \right )
\end{equation}
preserves the $P4/nmm$ space group symmetry since the latter is a scalar under all point group operations and is local in space ($I$ is a constant). It does however lock the orbital and spin parts of the conduction electrons together such that the two do not transform independently of each other under space group operations. This  forces us to change the irreducible representations of the space group to double valued representations \cite{Cracknell_book:1972}.
In general, the reflection $ P_{z} = \{ \sigma_{z} | 0 0 0\} $ is equivalent to

\noindent \begin{equation}
\label{Eq:Sigma_z}
\begin{matrix}
\sigma_z & = & C_{ 2z } \otimes I  \\
\end{matrix}
\end{equation}
\noindent where $C_{2z}$ corresponds to a rotation by $ \pi $ along a chosen $z$-axis and $I$ is the inversion. Since spinor states are invariant under the inversion \cite{Cracknell_book:1972}, the effect of $P_{z}$ is given by the Pauli matrix term
\begin{equation}
\label{Eq:Sigma_z_spinor}
D^{ ( j= \frac {1} {2} ) }( \sigma_z )=
\begin{pmatrix}
-i & & 0 \\
& & \\
0 & & i
\end{pmatrix}
\times
\begin{pmatrix}
1 & & 0 \\
& & \\
0 & & 1
\end{pmatrix}
=
\begin{pmatrix}
-i & & 0 \\
& & \\
0 & & i
\end{pmatrix}
= - i \sigma_{3}
\end{equation}

For arbitrary momentum $\pmb{k}$, one constructs the irreducible representations of the space group by determining the so-called group of the wave vector. That is the subgroup of the space group, with elements which either leave $\pmb{k}$ invariant or translate it by a reciprocal vector. By examining the effect of these elements on a set of properly chosen states, one can determine the group isomorphic to the group of the wave vector. In the case of double valued representations one must also consider, in addition to the operations in the single-valued case, those obtained by changing in sign of the states \cite{Cracknell_book:1972}.

Even with SOC terms, the glide reflection is still part of the group of the wave-vector. Because of the locking of the spin and orbital states, the glide reflection generates factors of $i$, consistent with Eq. \ref{Eq:Sigma_z_spinor}. By considering local states with and without spin and applying the operations that keep an arbitrary $\pmb{k}$ invariant one obtains a group isomorphic to the double valued group of $C_{1h}$. The irreducible representations of the latter are illustrated in Table \ref{Table:Double_rep}.

\noindent \begin{table}[htb]
\label{Table:Double_rep}
\caption{ The double-valued irreducible representations of the $ C_{1h} $ point group [ C.J. Bradley and A.P. Cracknell, \textit{Mathematical Theory of Symmetry in Solids} (Clarendon Press, Oxford 1972) ]. }
\begin{tabular}{r r r r r}
\\
\hline
\hline
$ C_{1h} $ & $ E $ & $ \sigma_{z} $ & $ \bar{E} $ & $ \bar{ \sigma}_{ z }  $ \\
\hline
$ \Gamma_{1} $ & $ 1 $ & $ 1 $ & $ 1 $ & $ 1 $ \\
$ \Gamma_{2} $ & $ 1 $ & $ -1 $ & $ 1 $ & $ -1 $ \\
$ \Gamma_{3} $ & $ 1 $ & $ i $ & $ -1 $ & $ -i $ \\
$ \Gamma_{4} $ & $ 1 $ & $ - i $ & $ -1 $ & $ i $ \\
\hline
\hline
\end{tabular}
\end{table}

The inclusion of the spin in the glide reflection will always generate the pure phases $\pm i$. We then conclude that the physical irreducible representations correspond to $\Gamma_{3}$ and $\Gamma_{4}$. This indicates that, as for the $P4/nmm$ case without SOC, for arbitrary momentum in an unfolded BZ, eigenstates of a general tight-binding Hamiltonian will also be eigenstates of the glide reflection symmetry. We can connect with the ansatz states in Eqs. \ref{Eq:Eigen_1_a}, \ref{Eq:Eigen_1_b} which coincide with Bloch states in a 1-Fe BZ. We remark that these are still eigenstates of the glide reflection operator but in the presence of SOC, we need to account for the transformation of the spins as well. We thus relabel

\noindent \begin{align}
C_{\pmb{k},E,\alpha, \uparrow} & \rightarrow C_{\pmb{k},\tilde{O},\alpha, \uparrow} \label{Eq:Corres_E_up}\\
C_{\pmb{k},E,\alpha, \downarrow} & \rightarrow C_{\pmb{k},\tilde{E},\alpha, \downarrow} \label{Eq:Corres_E_down} \\
C_{\pmb{k},O,\alpha, \uparrow} & \rightarrow C_{\pmb{k},\tilde{E},\alpha, \uparrow} \label{Eq:Corres_O_up} \\
C_{\pmb{k},O,\alpha, \downarrow} & \rightarrow C_{\pmb{k},\tilde{O},\alpha, \downarrow} \label{Eq:Corres_O_down}
\end{align}

\noindent where the $\tilde{E}, \tilde{O}$ refer to the sign in front of the $i e^{i \pmb{k} \cdot \pmb{\tau}}$ term under the glide transformation. Since the irreducble representations are also eigenstates of the glide, the Hamiltonian excluding the pairing terms can only connect $\tilde{E}, \tilde{E}$ or $\tilde{O}, \tilde{O}$ states. Note however that states $E,O$ states in the original (no SOC) labeling such as $C_{\pmb{k},E,\alpha, \uparrow}, C_{\pmb{k},O,\alpha, \uparrow}$ both belong to the $\tilde{E}$ irreducible representation of the space group with SOC. So in general, the space group symmetry allows the mixture of the $E,O$ states invalidating the unfolding procedure since the two will always correspond to states with shifted momenta $\pmb{k}$ and $\pmb{k}+ \pmb{Q}.$

Also note that the eigenstates of the Hamiltonian in this case are not degenerate along the Y line of the unfolded BZ \cite{Cvetkovic_Vafek:2013}, \cite{Cracknell_book:1972} (not counting Kramers degeneracy which is irrelevant here).  In general, there cannot be an analogous switching between the $\Gamma_3$ and $\Gamma_4$ representations at the edge of the BZ and one must conserve the ``parity" under the glide reflection as one crosses into adjacent zones. Therefore for finite SOC and for a general choice of tight-binding parameters,

\noindent \begin{equation}
\label{Eq:_NOT_BZ_translation}
C_{\pmb{k},E/O,\alpha}=C_{\pmb{k}+\pmb{Q}, O/E,\alpha}~~,
\end{equation}

\noindent with $\pm i$ eigenstates under the glide-reflection for $\Gamma_3$ and $\Gamma_4$ respectively.  This forces us to accept other eigenstates of the glide-reflection such as those in Eqs.~\ref{Eq:SOC_eigen_a},~\ref{Eq:SOC_eigen_b}.

The relation above also signals that we cannot choose the simple sublattice superposition states in (\ref{Eq:Eigen_1_a}),(\ref{Eq:Eigen_1_b}) and so we cannot connect with the effective extended momentum Fourier transforms in Eqs. \ref{Eq:Orb_Fourier_a}-\ref{Eq:Orb_Fourier_d} or the gauge transformed states in (\ref{Eq:Gauge_a}),(\ref{Eq:Gauge_b}). In a practical calculation where the 1-Fe unit cell assumption is of any use we implicitly carry on calculations using the familiar 1-Fe unit Fourier transform defined on $\pmb{\tilde{k}} \in UBZ$ and then fold to the physical 2-Fe unit BZ. But our arguments show that in doing so we are violating space group symmetry and as a consequence we have no guarantee that the results thus obtained coincide with those done directly in the unfolded zone.

As in the zero SOC case, we argue that it is insufficient to justify the validity or invalidity of the 1-Fe unit cell by invoking a general symmetry of the Hamiltonian, namely, the glide reflection. In the SOC case, the space group of the Hamiltonian does not change since the extra coupling does not violate any symmetry of the latter. Our arguments, through which we attempt to treat the glide reflection within its natural space group symmetry perspective, can be used to give a more precise prescription to its use and to illustrate how the formulation of the problem in an 1-Fe unit cell can fail in spite of the validity of the glide reflection as a symmetry of the Hamiltonian.

\subsection{The effects of the spin-orbit coupling on the normal-state bandstructure}
\label{Sec:SOC_numerics}

\begin{figure}[thb]
\includegraphics[width=80mm]{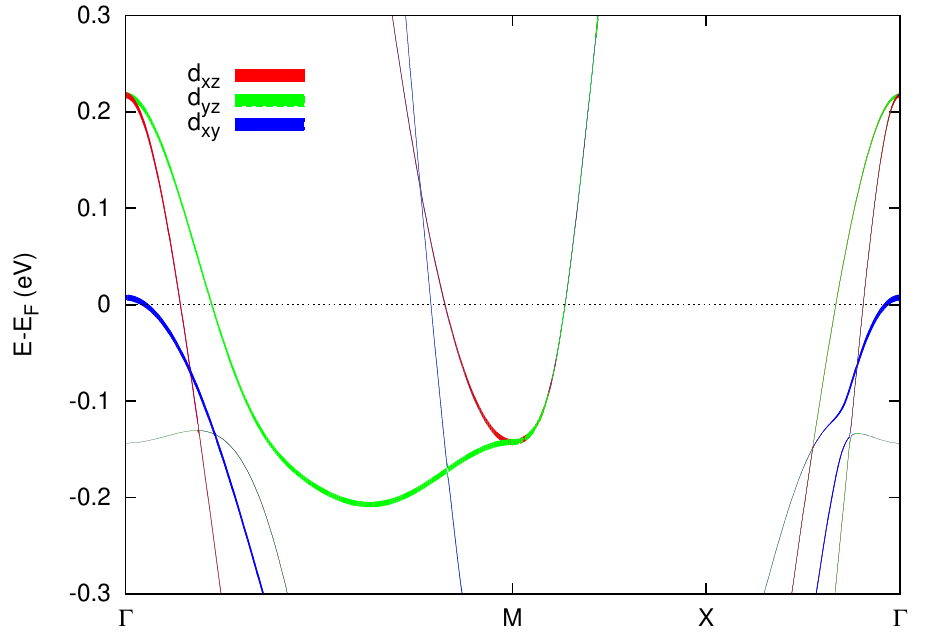}
\includegraphics[width=80mm]{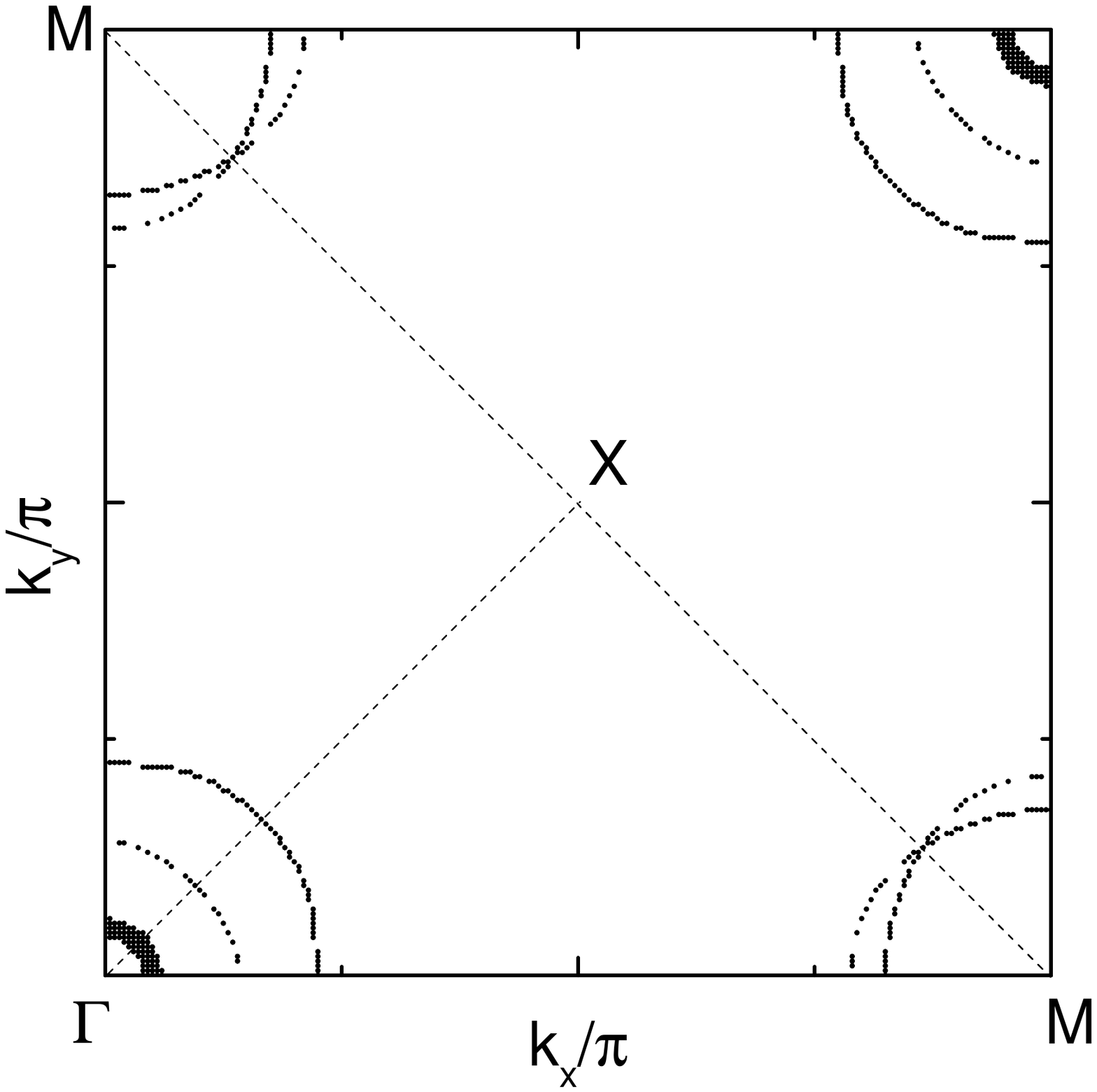}
\caption{ (a) Bandstructure of the five-orbital tight-binding model in the folded Brillouin zone (FBZ) without the spin-orbit coupling. (b) The corresponding Fermi surface in two quadrants of the FBZ.}
\label{Fig:Bands1}
\end{figure}

\begin{figure}[thb]
\includegraphics[width=80mm]{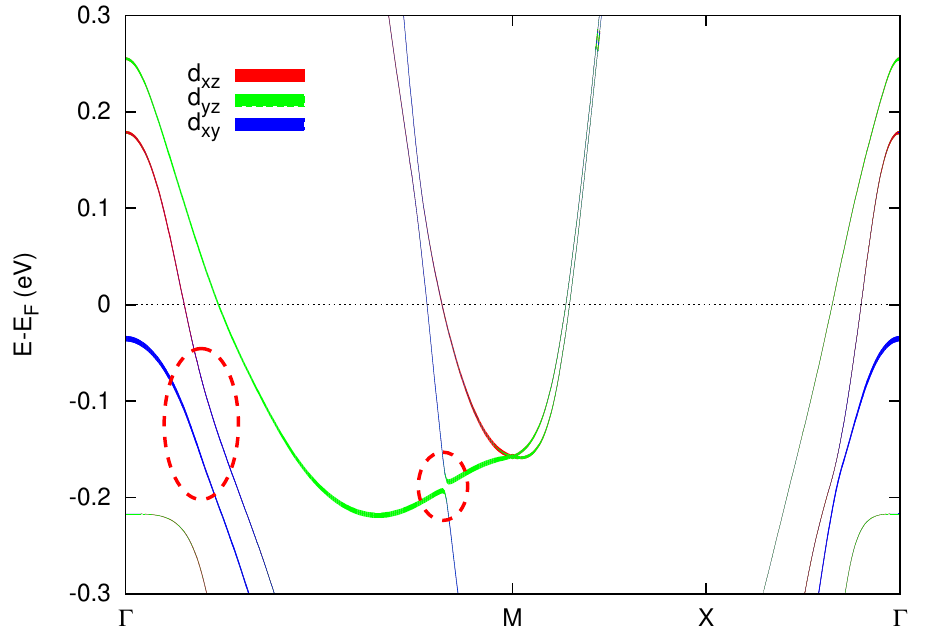}
\includegraphics[width=80mm]{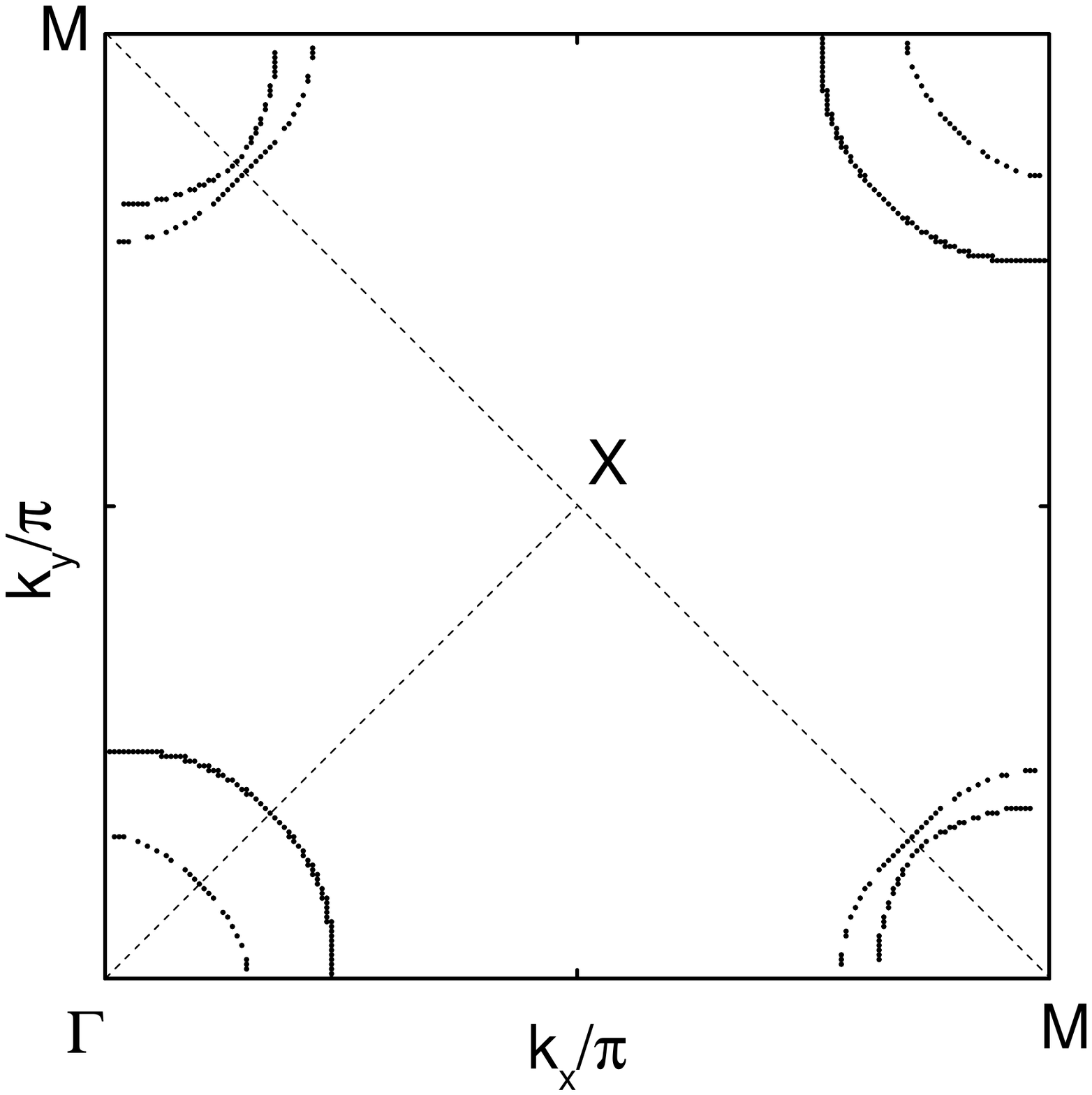}
\caption{(a) Bandstructure of the five-orbital tight-binding model in the folded Brillouin zone (FBZ) with a local spin-orbit coupling $\lambda_{\rm{SO}}=0.05$ eV. (b) The corresponding Fermi surface in two quadrants of the FBZ.}
\label{Fig:Bands2}
\end{figure}

In light of the discussion on the folding procedure, in this section we study the effects of the spin-orbit coupling on the bandstructure of the normal state. We consider the following Hamiltonian: $H=H_{\rm{TB}} + H_{\rm{SO}}$, where
\begin{eqnarray}
 H_{\rm{TB}} = \sum_{\mathbf{k}\alpha\beta,\sigma} \epsilon_{\alpha\beta} (\mathbf{k}) C^\dagger_{\mathbf{k}\alpha\sigma} C_{\mathbf{k}\beta\sigma},
\end{eqnarray}
is a five-orbital tight-binding model for the parent compound of iron pnictides.
Here $\epsilon_{\alpha\beta} (\mathbf{k})$ are tight-binding parameters, which we adopted from Ref.~\onlinecite{Graser:2009}. $H_{\rm{SO}} = \lambda_{\rm{SO}} \sum_{i} \mathbf{L}_i\cdot\mathbf{S}_i$, refers to a local spin-orbit coupling term. As we discussed, this term does not break any symmetry of the lattice, but couples the spatial and spin part of the single-particle wave function.

According to Sec.~\ref{Subsection:Folding_wo_SOC}, the tight-binding Hamiltonian defined by the pseudo-crystal momentum $\pmb{\tilde{k}}$ in the UBZ can be transformed to the physical momentum in FBZ via the folding procedure. This allows us to study the effects of the spin-orbit coupling by comparing the bandstructures without and with a spin-orbit coupling in the FBZ. In the absence of the spin-orbit coupling, the bandstructure of the tight-binding model is shown in panels (a) and (b) of Fig.~\ref{Fig:Bands1}; it is in agreement with that from {\it ab~initio} calculations using the density functional theory~\cite{Graser:2009}. It is clearly seen that the bands are doubly degenerate along the boundary of the FBZ (Y line from the M to the X point), and as a consequence, the two elliptical electron pockets cross at a point along this direction. As we have emphasized in Sec.~\ref{Subsection:Folding_wo_SOC}, this double degeneracy guarantees the successful folding procedure: the wave function can switch representations under the glide when crossing the FBZ boundary, and hence one can define a pseudo-crystal
 momentum according to the parity of the wave function under the glide. This ensures the equivalence between the models defined in the FBZ and UBZ. When a spin-orbit coupling is turned on, as shown in Fig.~\ref{Fig:Bands2} (a) and (b), the double degeneracy along the FBZ boundary (M-X direction) is lifted. As we discussed, this invalidates the unfolding procedure since the wave functions have to conserve the parity under glide across the FBZ boundary. As a result of the lifted degeneracy, the two electron pockets no longer cross, but a gap opens along the X-M direction.

Another feature of the bandstructure in the presence of spin-orbit coupling is the opening of hybridization gaps between the $d_{xy}$ and $d_{xz/yz}$ bands along the $\Gamma$-M direction as shown within the dashed ellipses in Fig.~\ref{Fig:Bands2}. In absence of spin-orbit coupling, these bands simply cross, without opening a gap because they have different parity (pseudo-crystal momenta) and can not mix. But with a finite spin-orbit coupling, the $C_{\mathbf{k},xz/yz,\uparrow}$ and $C_{\mathbf{k+Q},xy,\downarrow}$ have the same parity and they can hybridize via opening a gap. Recently, a hybridization gap between the $d_{xz/yz}$ and $d_{xy}$ bands along the $\Gamma$-M direction has been observed in ARPES measurements on several iron-based compounds~\cite{Yi:2011,Liu:2015}. It would be interesting to compare the experimental results with theoretical ones as this may provide valuable information about the strength of spin-orbit coupling in these systems. The spin-orbit couplin
 g also mixes the $d_{xz}$ and $d_{yz}$ orbitals, and lifts the double degeneracy between the $d_{xz}$ and $d_{yz}$ orbitals at $\Gamma$ point. But a local spin-orbit coupling does not lift the degeneracy between the $d_{xz}$ and $d_{xy}$ orbitals at the bottom of the electron bands at M point since it does not break the four-fold rotational symmetry of the P4/nmm group.

\subsection{The pairing interactions in the presence of spin-orbit coupling}
\label{Sec:Pairing_SOC}

In Sec. \ref{Sec:Space_group_with_SOC} we showed that the correspondence between states defined on 
a 1-Fe BZ and those belonging to the irreducible representations of the space group in the presence 
of a SOC breaks down even when there are no pairing interactions present. For consistency, 
here we analyze the direct effect of a local spin-orbit coupling term (Eq. \ref{Eq:SOC_term}) 
on the pairing interactions of a $t-J$ model.

The projective nature of this model \cite{Goswami:2009} excludes non-singlet pairing terms. However, 
the SOC does not conserve the electronic spin quantum number and in this case a $t-J$ model 
must include ad-hoc triplet pairing terms. In general, the resulting pairing must include terms
 which are odd under inversion. Based on the discussion at the beginning of Sec.~\ref{Sec:Effects_pairing}, 
 we expect that the inversion-odd pairing, 
 in the spin triplet case,
will also be odd under the glide-reflection and thus correspond to finite-momentum pairs in the UBZ. We note that, as in the case for pairing without SOC, we restrict ourselves to one-dimensional representations of the point group, for which the above is correct. In the case of two-dimensional representations, the opposite holds, with inversion-even functions actually corresponding to glide-reflection odd states.

At mean-field level, the pairing can still be written in the form of Eq. \ref{Eq:MF_term} except for the spin structure allowing both same-spin and opposite spin pairing. More specifically, we use the correspondence in Eqs. \ref{Eq:Corres_E_up}-\ref{Eq:Corres_O_down} between the $E,O$ labels without SOC and the $\tilde{E}, \tilde{O}$ ones with SOC turned on to re-write the pairing in Eq. \ref{Eq:Pairing_MF} in terms of the latter.
 As a consequence, the real-space pairing functions
 $\Delta^{A}_{\pmb{e},\alpha\beta}, \Delta^{B}_{\pmb{e},\alpha\beta}$ (Eqs. \ref{Eq:Delta_A}, \ref{Eq:Delta_B})
 acquire a spin-index dependence. The previous argument (without SOC) relied on the transformation
 properties of the latter term under the glide reflection. We can apply the same procedure in real-space
 allowing the transformation of the spins. For singlet pairing the $\uparrow, \downarrow$ combination
  always generate products of $i, -i$, ensuring that $\Delta^{A,B}_{\alpha, \beta}$ have exactly the same properties as before. Applying the transformation (\ref{Eq:Corres_E_up})-(\ref{Eq:Corres_O_down}) in reverse, we note that singlet pairing can only have diagonal $E,E$ and
 $O,O$ pairing (original labeling) and so does not introduce any finite momentum Cooper pairs.

The situation is different when we allow triplet pairing. The same-spin terms always generate a minus
sign under the glide reflection. When $\alpha, \beta$ have the same parity this means
$\Delta^{A}_{\pmb{e},\alpha\beta}=-\Delta^{B}_{\pmb{e},\alpha\beta}$. In the original labeling,
the only terms that survive are the off-diagonal $E,O$. Similarly, when $\alpha, \beta$ have different
parity $\Delta^{A}_{\pmb{e},\alpha\beta}=\Delta^{B}_{\pmb{e},\alpha\beta}$ but the $(-1)$ terms coming
from the orbital parity  guarantee that we again obtain only off diagonal $E,O$ pairs.
Now however, the $E,O$ terms will map to $\pmb{\tilde{k}},\pmb{\tilde{k}+ Q}$ in the UBZ producing
 finite-momentum Cooper pairs and invalidating the unfolding. The spin-symmetric combinations allow forms 
like those for singlet pairing and thus do not introduce any new terms.

\section{Discussions}
\label{Sec:Discussion}
In this section we elaborate on some further issues related to the space group symmetry and BZ folding we detailed above.
\subsection{Effects of three-dimensionality vs. spin-orbit coupling}
Real materials with a $P4/nmm$ space group have a 3D structure ensuring the bands are always dispersive, albeit weakly, along the $k_z$ direction. The group of the wave-vector for arbitrary $k_{z}$ cannot contain the glide reflection since the latter connects the generally inequivalent $k_{z}, -k_{z}$ components. For $k_{z}=0$, the folding can still work as we illustrate below. The dispersion along the $z$-direction can be accounted for by generalizing the 2D ansatze (\ref{Eq:Eigen_1_a}), (\ref{Eq:Eigen_1_b}) to

\noindent \begin{align}
C_{\pmb{k},k_{z},E,\alpha} = & \frac{1}{\sqrt{2}} \sum_{R_{z}}  e^{ik_{z}R_{z}} \left[ C_{\pmb{k},A,\alpha, R_{z}} + (-1)^{\alpha}  C_{\pmb{k},B,\alpha, R_{z}} \right], \label{Eq:Eigen_1_a_z} \\
C_{\pmb{k},k_{z},O,\alpha} = & \frac{1}{\sqrt{2}} \sum_{R_{z}}  e^{ik_{z}R_{z}} \left[ C_{\pmb{k},A,\alpha,  R_{z}} - (-1)^{\alpha}  C_{\pmb{k},B,\alpha,  R_{z}} \right], \label{Eq:Eigen_1_b_z}
\end{align}

\noindent which for general $k_{z}$ are not eigenstates of $T_{\tau} P_{z}$ and we keep the $E,O$ indices for labeling purposes. Here, $\pmb{k}$ still refers to a purely 2D wave vector. Since for finite $k_{z}$, these states do not necessarily coincide with the irreducible representations of the space group, the Hamiltonian will contain terms mixing $E,O$ indices. For $k_{z}=0$ however, the glide does not affect the Bloch momentum label but it maps $R_{z} \rightarrow -R_{z}$. Due to the equivalence of the different planes, $C_{\pmb{k},A/B,\alpha, R_{z}}=C_{\pmb{k},A/B,\alpha, -R_{z}}$ since the two Wannier states are related by a proper translation. This means in the $k_{z}=0$ plane the states (\ref{Eq:Eigen_1_a_z}), (\ref{Eq:Eigen_1_b_z}) still coincide with the irreducible representations of the space group and the previous arguments still apply. Therefore, states with two-dimensional crystal momenta can still be mapped onto an effective 1-Fe BZ even when the dispersion along the z-axis is turned on.

\subsection{Space group symmetry and parity of pairing}
Recent studies have paid particular attention to the consequences specific to the 2-Fe unit cell
such as the use of an additional isospin quantum number \cite{Ong_Coleman:2013}
on the superconducting pairing. But some important issues are still unclear.
For example, pairing channels with very different symmetries have been proposed
theoretically, \cite{Ong_Coleman:2013,Hu:2013} although the target compounds have the same spatial
symmetry. As we discussed earlier in our paper, the space group
 already imposes strong constraints on the parity of symmetry-compatible pairing.
 This provides a means to check whether a pairing channel is  allowed by
 the space group symmetry of the system. Here we examine this issue
 for two recently proposed pairing channels, the TAO pairing~\cite{Ong_Coleman:2013}
 and the $\eta$-pairing~\cite{Hu:2013}. As before, we focus on spin-singlet pairings in one-dimensional representations henceforth.

\subsubsection{TAO pairing}
Discussed in Ref.~\onlinecite{Ong_Coleman:2013}, the TAO pairing refers to a spin singlet, orbital triplet $A_{1g}$ $d_{x^2-y^2}+id_{xy}$ pairing in the 2-Fe BZ. It is easy to check that this pairing channel has an even parity and is compatible with the $P4/nmm$ space symmetry. From Sec.~\ref{Sec:Effects_pairing} we know that for a 2D dispersion of the conduction electrons and in the absence of spin-orbit coupling,
any pairing defined on the 2-Fe unit cell FBZ which respects the space group symmetry must have an equivalent, albeit given by a different linear combination of channels, $(\pmb{\tilde{k}},-\pmb{\tilde{k}})$ pairing
 in the 1-Fe unit cell UBZ. The aforementioned TAO pairing defined in the 2-Fe BZ $(\pmb{k},-\pmb{k})$ is equivalent to one defined in the 1-Fe UBZ and does not incorporate any particular properties of the 2-Fe BZ not captured by the former. A possible advantage of the direct 2-Fe unit cell formulation might consist in expounding physical features which might be harder to illustrate in an equivalent 1-Fe unit cell picture.

\subsubsection{$\eta$-pairing}
Recent  discussions have also 
considered
the $\eta$-pairing in the iron-based 
superconductors ~\cite{Hu:2013,Hu_Hao_Wu:2013,Lin_Ku:2014,Wang_Maier:2014}.
In its original proposal,~\cite{Hu:2013} 
the $\eta$-pairing refers to
a singlet pairing of two electrons with pseudo-crystal momenta $\pmb{\tilde{k}}$ and $-\pmb{\tilde{k}+Q}$, 
respectively. This pairing has nonzero total pseudo-crystal momentum  in the 1-Fe UBZ, and the momentum 
dependent part of the wavefunction has odd parity under inversion.
This proposal was based on the observation that the inversion center in an Fe plane lies half-way in between two (inequivalent) sites. In a real space basis, the inversion operation interchanges the two positions on the different sublattices. In a simpler case where the two lattices are equivalent, the antisymmetry under exchange of the pairing wavefunction forces the spatial ($\pmb{k}$-dependent) part to be even under inversion. The existence of the two inequivalent sites opens up the possibility of odd parity momentum dependence since the overall spatial part could be described as a direct product of a purely $\pmb{k}$-dependent part (in the FBZ, 2-Fe unit cell description) 
and a pseudo-spin matrix which captures the effect of the different sublattices. The purely $\pmb{k}$-dependent 
part can have odd parity under inversion as long as the remaining degrees of freedom compensate with a minus 
sign such that the total inversion parity is still even. Although not explicitly stated in Ref.~\onlinecite{Hu:2013}, 
such a state could correspond to a $E,O$ pseudo-spin singlet. Upon unfolding, the combination 
will generate the finite momentum $(\pmb{\tilde{k}},-\pmb{\tilde{k}+Q})$ pairing. 
As already shown in Sec.~\ref{Sec:Effects_pairing},
for states of definite parity under inversion, or equivalently, of definite parity under the glide-reflection,
the pairing can be written  in general as a linear combination of terms with and without $E,O$ mixing together 
with their respective momentum-dependent parts. In addition, if we do not break the glide-reflection symmetry, 
the total pairing must be even. In a purely two dimensional BZ, the glide-reflection cannot change any of the 
momenta of the pair, such that terms with $E,O$ mixing change sign under the operation while those without 
do not. Indeed, as detailed in our Sec.~\ref{Sec:Effects_pairing} , all mixing terms must vanish, 
ensuring that upon unfolding no finite momentum pairs can be generated. 
We stress that this conclusion holds whenever inversion is not spontaneously broken for a two dimensional BZ.

In recent studies on the $\eta$-pairing, there also seems to be confusion in the definition 
for this type of pairing in the literature which, we believe, is associated with  different subsequent
definitions of the 1-Fe BZ. While in the original work the $\eta$-pairing 
referred to a $(\pmb{\tilde{k}},-\pmb{\tilde{k}+Q})$ state, in more recent studies, it is associated with
a $(\pmb{\slashed{k}},-\pmb{\slashed{k}+Q})$ pairing in the so-called
 ``physical extended BZ"~\cite{Lin_Ku:2014,Wang_Maier:2014} given by 
 a mapping similar to the one in Eqs. \ref{Eq:Orb_Fourier_a}-\ref{Eq:Conventioanal_1}. 
 As in the original proposal \cite{Hu:2013}, Ref. \onlinecite{Lin_Ku:2014} 
 alludes to the inequivalence of the two sublattice sites but does not attempt to explicitly 
 consider eigenstates of the glide reflection operation. 
 Rather, the authors classify 2D states both in the "physical" ($\pmb{\slashed{k}}$ in our convention) 
 and unfolded ($\pmb{\tilde{k}}$) representations according to the parity under the reflection about the z-plane. 
 In the physical representation, an alternating minus sign in the hopping between different orbital parity Wannier 
 states is "absorbed" into the definition of the quasi-crystal momentum, resulting in a shift by $\pmb{Q}$ 
 between the even and odd reflection parity states, as detailed in our Sec. \ref{Subsection:Folding_wo_SOC}. 
 The authors argue that a zero mometum pair in the UBZ ($\pmb{\tilde{k}}, - \pmb{\tilde{k}})$ 
 or arbitrary parity under \emph{the z-reflection alone} must decompose into a linear combination of even/even, 
 odd/odd and even/odd terms which
  correspond to $(\pmb{\slashed{k}}+ \pmb{Q}, 
  - \pmb{\slashed{k}}-\pmb{Q})$, $(\pmb{\slashed{k}}, - \pmb{\slashed{k}})$ 
  and $(\pmb{\slashed{k}}+ \pmb{Q}, - \pmb{\slashed{k}} )$ respectively in the physical BZ. 
  From this, they seem to explain inconsistencies in the spectral weights 
  in the superconducting state between calculations done on a 1-Fe/UBZ ($\pmb{\tilde{k}}$) 
  and subsequently folded down and ARPES experiments among others whose results are naturally 
 obtained in a 2-Fe/FBZ.

In the context of a glide reflection symmetry, the two constructions referring to 
a finite momentum $(\pmb{\tilde{k}},-\pmb{\tilde{k}+Q})$ in the UBZ~\cite{Hu:2013} and 
in the "physical" $(\pmb{\slashed{k}},-\pmb{\slashed{k}+Q})$ ~\cite{Lin_Ku:2014,Wang_Maier:2014} 
are different. The $(\pmb{\tilde{k}},-\pmb{\tilde{k}+Q})$ pairing has an odd parity under glide reflection, 
while the $(\pmb{\slashed{k}},-\pmb{\slashed{k}+Q})$ -pairing is parity even under the glide. 
To see this, consider the latter ("physical" representation) case which corresponds to an odd/even 
parity combinations under \emph{the pure reflection}. In our language, this corresponds to states 
given in Eqs.~\ref{Eq:Orb_Fourier_a}-\ref{Eq:Orb_Fourier_d} with $\alpha$ odd and even respectively. 
It is not difficult to see that upon converting to the "true" 2-Fe/FBZ ($\pmb{k}$) this term corresponds to ($\pmb{k}, -\pmb{k}), O/O$  under the \emph{glide-reflection} pairs. 
To the best of our knowledge, this point has not been clarified before.
According to our analysis, the only symmetry-allowed pairing is the even glide parity one, \emph{i.e.}, the $(\pmb{\tilde{k}},-\pmb{\tilde{k}})$ pairing (or equivalently, the $(\pmb{\slashed{k}},-\pmb{\slashed{k}+Q})$ pairing).
It corresponds to normal zero-momentum pairing in both the 2-Fe FBZ ($\pmb{k}$) or the 1-Fe UBZ in pseudo-crystal momentum space $ (\pmb{\tilde{k}}) $. More precisely, the three formulations alluded to above are equivalent and we see no reason why 1-Fe unit cell/UBZ calculations which are folded down could not \emph{a priori} capture experimental results. This correspondence naturally explains why the superconducting gap functions obtained from calculations in the 2-Fe FBZ are identical to the previous results in the 1-Fe UBZ~\cite{Wang_Maier:2014}. Within our approach, we have also shown
 that the odd glide parity $(\pmb{\tilde{k}},-\pmb{\tilde{k}+Q})$ pairing is not allowed by symmetry.

\section{Conclusions}
\label{Sec:Conclusion}

The glide reflection symmetry is valid for states of arbitrary momentum and without spin-orbit coupling there is a minimal double degeneracy all along the 2-Fe unit cell BZ edge. This ensures that a tight-binding Hamiltonian can be determined using an unfolded BZ corresponding to a 1-Fe unit cell. By contrast, although the glide symmetry
still holds for arbitrary momentum when spin-orbit coupling is turned on, the latter mixes states corresponding to different pseudo-momenta in the unfolded BZ and lifts the degeneracy along the $Y$ line, forbidding the general use of the same unfolding procedure. These conclusions are consistent with bandstructure calculations with and without spin-orbit coupling which show the lack of and the presence of this hybridization.

We also conclude that for a $t-J$ type Hamiltonian without spin-orbit coupling the results obtained directly
from a 1-Fe unit cell should coincide with those from 2-Fe unit cell, the two being related
by the validity of the unfolding procedure. This applies to the TAO pairing~\cite{Ong_Coleman:2013}:
Though proposed in the 2-Fe BZ, it is equivalent to a $d_{x^2-y^2}+id_{xy}$ pairing with both intra- and
inter-orbital contributions. One more remark is that this equivalence does not hold when a spin-orbit coupling
term is included.

Another conclusion from our symmetry analysis is that the pairing channel compatible to the $P4/nmm$ space group symmetry must have an even parity (once again,we focus on spin-singlet pairings in one-dimensional representations).
With this criterion, the $\eta$-pairing with $(\pmb{\tilde{k}},-\pmb{\tilde{k}}+Q)$, originally proposed in Ref.~\cite{Hu:2013} is not symmetry-allowed since it is parity odd.
The $\eta$-pairing discussed in most recent works~\cite{Lin_Ku:2014,Wang_Maier:2014}, on the other hand,
refers to a $(\pmb{\slashed{k}},-\pmb{\slashed{k}+Q})$ pairing, which corresponds to a total momentum
zero $(\pmb{\tilde{k}},-\pmb{\tilde{k}})$ pairing with an even parity, and is thus compatible with
the $P4/nmm$ space group symmetry.

Our analysis for the folding within a $P4/nmm$ space group symmetry can serve as a comparison point for a similar discussion in the more involved $I4/mmm$ case. As we stressed throughout the text, the validity of the glide-symmetry, together with the fortuitous double degeneracy along the edge of the FBZ guarantee the success of the folding for $k_{z}=0$ and no SOC. Since none of these appear to be valid for $I4/mmm$,
 the folding will probably not work. A rigorous analysis of this latter case is reserved for a future publication.

 {\it Acknowledgements.~}
This work has been supported by the NSF Grant No. DMR-1309531 and the Robert A.\ Welch Foundation
Grant No.\ C-1411 (E.M.N. \& Q.S.).
R.Y. was partially supported by the National Science Foundation of China
Grant number 11374361, and the Fundamental Research Funds for the
Central Universities and the Research Funds of Renmin University
of China.
All of us acknowledge the support provided in part by the
NSF Grant  No. NSF PHY11-25915 at KITP, UCSB, for our participation in the Fall 2014
program on ``Magnetism, Bad Metals and Superconductivity: Iron Pnictides and Beyond".
Q.S.\ also acknowledges the hospitality
of the Institute of Physics of Chinese Academy of Sciences.

\appendix

\section{Eigenstates of the glide operation}
\label{Sec:Appendix_A}

The irreducible representations of the $P_{4/nmm}$ space group for general $\pmb{k} \in FBZ$ and for the special loci $\Gamma$, $\Delta$, $\Sigma$ and $M$ \cite{Cvetkovic_Vafek:2013} transform as

\noindent \begin{align}
\label{Eq:Glide}
(T_{ \pmb { \tau } } P_z) C_{\pmb{k}\alpha} = e^{ i \pmb{k} \pmb{\tau}} \lambda C_{\pmb{k}\alpha},
\end{align}

\noindent where $\lambda= \pm 1$. The above form is not the case for $X$ and $Y$. However, since the Hamiltonian must evolve continously with $\pmb{k}$ it is clear that we can form irreducible representations at the two above mentioned loci by taking linear combinations of even ($E$) only or odd only ($O$) such that our arguments are not affected.

We can derive the general form of states which transform according to (\ref{Eq:Glide}) from a general superposition of operators defined on each sublattice

\noindent \begin{align}
\label{Eq:general_form}
C_{\pmb{k}\alpha}= & \frac{1}{\sqrt{N}} \left[\sum_{\pmb{R_A}}e^{i\pmb{k} \cdot \pmb{R_A}} C^{(A)}_{\pmb{R_A}\alpha} + \sum_{\pmb{R_B}}   e^{i\theta_{\pmb{k}}}e^{i\pmb{k}\pmb{R_{B}}}C^{(B)}_{\pmb{R_B} \alpha} \right] \notag \\
 = & \frac{1}{\sqrt{N}} \sum_{\pmb{R_A}} e^{i\pmb{k} \cdot \pmb{R_A}}  \left[ C^{(A)}_{\pmb{R_A}\alpha} + e^{i\theta_{\pmb{k}}}e^{i\pmb{k} \pmb{\tau} }C^{(B)}_{\pmb{R_A}+\pmb{\tau} \alpha} \right]
\end{align}

\noindent where $\pmb{R}_{A}$, $\pmb{R}_{B}$ are summations over the position vectors of the respective sublattices, and $\alpha$ stands for the parity of the local degrees of freedom under a pure reflection $\sigma_z$. We stress that, in the most general case, the completely local states can be different ($C^{(A)}_{\pmb{r} \alpha} \neq C^{(B)}_{\pmb{r}\alpha}$) due to the physical inequivalence of the two sublattice sites. Although in principle the linear combinations of the two sublattice Bloch states can have arbitrary complex coefficients, the form chosen above is sufficient for our purposes.

Applying the glide reflection to the trial state (\ref{Eq:general_form}) and imposing (\ref{Eq:general_form}) we get

\noindent
\begin{widetext}
\begin{align}
\label{Eq:Transformed_state}
(T_{ \pmb { \tau } } P_z) C_{\pmb{k}\alpha}=  \lambda e^{ i \pmb{k} \pmb{\tau}} \frac{1}{\sqrt{N}} \sum_{R_A} e^{i\pmb{k} \cdot \pmb{R_A}} \left[  (-1)^\alpha C^{(A)}_{\pmb{R_A}+\pmb{\tau}\alpha} + e^{i\tilde{\theta_{\pmb{k}}}}e^{i\pmb{k}\pmb{\tau}} (-1)^\alpha C^{(B)}_{\pmb{R_A} +2\pmb{\tau} \alpha} \right]
\end{align}
\end{widetext}

\noindent such that all position vectors get shifted by the fractional translation $\pmb{\tau}$, the local states generate the parity term $\alpha$, and the phase $\theta_{ \pmb{ k } } \rightarrow \tilde{ \theta_{\pmb{k}} } $  in general. Comparing (\ref{Eq:general_form}) and (\ref{Eq:Transformed_state}) we see there are a number of possibilities. In general, condition (\ref{Eq:Transformed_state}) cannot determine all the unknowns i.e. the phase factor and the relation between the the two displaced local states. We can connect with a folding procedure by making some assumptions regarding the phase and letting the above condition determine the local states.

A first possibility corresponds to taking $\lambda=1$ in (\ref{Eq:Transformed_state})

\noindent \begin{align}
\label{Eq:1_poss_1}
 C^{(A)}_{\pmb{R_A}\alpha}= &  (-1)^\alpha C^{(A)}_{\pmb{R_A}+\pmb{\tau}\alpha} \\
C^{(B)}_{\pmb{R_A} + \pmb{\tau}\alpha}= &  (-1)^\alpha C^{(B)}_{\pmb{R_A}+2\pmb{\tau}\alpha} \\
\theta_{ \pmb{ k } } = & \tilde{ \theta_{\pmb{k}} }.
\end{align}

\noindent For $\lambda=-1$ we can have

\noindent \begin{align}
\label{Eq:1_poss_2}
 C^{(A)}_{\pmb{R_A}\alpha}= & - (-1)^{\alpha} C^{(A)}_{\pmb{R_A}+\pmb{\tau}\alpha} \\
C^{(B)}_{\pmb{R_A} + \pmb{\tau}\alpha}= & - (-1)^{\alpha} C^{(B)}_{\pmb{R_A}+2\pmb{\tau}\alpha} \\
\theta_{ \pmb{ k } } = & \tilde{ \theta_{\pmb{k}} }.
\end{align}

\noindent and  we set $\theta_{ \pmb{ k } }=0$.  The conditions above are simply the transformation properties of the local states under a simple 1-Fe unit cell translation. If we choose the states at corresponding to $C_{A}$ and $C_{B}$ to have the same functional form, we can construct two distinct eigenstates of the glide by for arbitrary $\alpha$ :

\noindent \begin{align}
C_{\pmb{k},E,\alpha}= & \frac{1}{\sqrt{N}} \sum_{\pmb{R_A}} e^{i\pmb{k} \cdot \pmb{R_A}}  \left[ C_{\pmb{R_A}\alpha 1} + (-1)^{\alpha} e^{i\pmb{k} \pmb{\tau} }C_{\pmb{R_A}+\pmb{\tau} \alpha } \right] \label{Eq:App_state_a} \\
C_{\pmb{k},O,\alpha2}= & \frac{1}{\sqrt{N}} \sum_{\pmb{R_A}} e^{i\pmb{k} \cdot \pmb{R_A}}  \left[ C_{\pmb{R_A}\alpha} -(-1)^{\alpha} e^{i\pmb{k} \pmb{\tau} }C_{\pmb{R_A}+\pmb{\tau} \alpha} \right]  \label{Eq:App_state_b},
\end{align}

\noindent  where we omitted the $A$,$B$ superscripts which are now irrelevant. Eqs. \ref{Eq:App_state_a}  and \ref{Eq:App_state_b} are the operators in (\ref{Eq:Eigen_1_a}) and (\ref{Eq:Eigen_1_b}).

Note that in addition to the above states which allowed the unfolding we can also choose eigenstates of the glide reflection by imposing

\noindent \begin{align}
 C^{(A)}_{\pmb{R_A}\alpha}= & \pm (-1)^{\alpha} e^{i\tilde{\theta_{\pmb{k}}}}e^{i\pmb{k}\pmb{\tau}} C^{(B)}_{\pmb{R_A}+2\pmb{\tau}\alpha} \\
e^{i\theta_{\pmb{k}}}e^{i\pmb{k} \pmb{\tau} }C^{(B)}_{\pmb{R_A}+\pmb{\tau} \alpha}= & \pm (-1)^{\alpha} C^{(A)}_{\pmb{R_A}+\pmb{\tau}\alpha}.
\end{align}

\noindent We can choose the phase factor $e^{i \theta_{ \pmb{ k } }} = e^{i \tilde{ \theta_{\pmb{k}} }}= e^{-i\pmb{k\tau}}$ and disregard the $A,B$ distinction as in the first choice. The two sublattices are now related by a $\pmb{k}$-dependent phase rather than a simple sign. The initial ansatz (\ref{Eq:general_form}) becomes

\noindent \begin{align}
C_{\pmb{k},E, \alpha}= & \sqrt{\frac{1}{N}} \sum_{\pmb{R_A}} e^{i\pmb{k} \cdot \pmb{R_A}}  \left[ C_{\pmb{R_A}\alpha} +(-1)^{\alpha} C_{\pmb{R_A}+\pmb{\tau} \alpha} \right] \label{Eq:SOC_eigen_a} \\
C_{\pmb{k},O,\alpha}= & \sqrt{\frac{1}{N}} \sum_{\pmb{R_A}} e^{i\pmb{k} \cdot \pmb{R_A}}  \left[ C_{\pmb{R_A}\alpha} - (-1)^{\alpha} C_{\pmb{R_A}+\pmb{\tau} \alpha} \right] \label{Eq:SOC_eigen_b}
\end{align}

\noindent which clearly do not correspond to a 1-Fe unit cell.

\section{The Hamiltonian in the physical extended momentum basis}
\label{Sec:Appendix_B}

Direct substitution of the definitions (\ref{Eq:Orb_Fourier_a})-(\ref{Eq:Orb_Fourier_d}) into the 2-Fe BZ Hamiltonian (\ref{Eq:2_Fe_Hamiltonian}) gives

\begin{widetext} \begin{align}
H_{TB}= & \sum_{\pmb{k} \in FBZ} \Bigg[ \sum_{ee} \left( \epsilon^{E}_{\alpha \beta}(\pmb{k}) C^{\dagger}_{\pmb{k}, \alpha} C_{\pmb{k}, \beta} + \epsilon^{O}_{\alpha \beta}(\pmb{k}) C^{\dagger}_{\pmb{k}+ \pmb{Q}, \alpha} C_{\pmb{k}+\pmb{Q}, \beta} \right) + \sum_{oo} \left( \epsilon^{E}_{\alpha \beta}(\pmb{k}) C^{\dagger}_{\pmb{k}+\pmb{Q}, \alpha} C_{\pmb{k}+\pmb{Q}, \beta} + \epsilon^{O}_{\alpha \beta}(\pmb{k}) C^{\dagger}_{\pmb{k}, \alpha} C_{\pmb{k}, \beta} \right) \nonumber \\
& + \sum_{eo} \left( \epsilon^{E}_{\alpha \beta}(\pmb{k}) C^{\dagger}_{\pmb{k}, \alpha} C_{\pmb{k}+\pmb{Q}, \beta} + \epsilon^{O}_{\alpha \beta}(\pmb{k}) C^{\dagger}_{\pmb{k}+\pmb{Q}, \alpha} C_{\pmb{k}, \beta} \right) + \sum_{oe} \left( \epsilon^{E}_{\alpha \beta}(\pmb{k}) C^{\dagger}_{\pmb{k}+\pmb{Q}, \alpha} C_{\pmb{k}, \beta} + \epsilon^{O}_{\alpha \beta}(\pmb{k}) C^{\dagger}_{\pmb{k}, \alpha} C_{\pmb{k}+\pmb{Q}, \beta} \right) \Bigg].
\end{align}
\end{widetext}

\noindent We can use the fact that $\epsilon^{E(O)}_{\alpha \beta}(\pmb{k}+\pmb{Q})= \epsilon^{O(E)}_{\alpha \beta}(\pmb{k})$ to rewrite the first two terms as a total sum over $\pmb{\slashed{k}}$ over the unfolded BZ. We can also show that the third term can be expressed as a sum over $\pmb{\slashed{k}}$. Specifically,

\noindent \begin{widetext}
\noindent \begin{align}
\sum_{\pmb{\slashed{k}} \in UBZ} \epsilon^{E}_{\alpha \beta} (\pmb{\slashed{k}}) C^{\dagger}_{\pmb{\slashed{k}}, \alpha \beta}C_{\pmb{\slashed{k}}+\pmb{Q}, \alpha \beta}= & \sum_{\pmb{\slashed{k}} \in FBZ} \epsilon^{E}_{\alpha \beta} (\pmb{\slashed{k}}) C^{\dagger}_{\pmb{\slashed{k}}, \alpha \beta}C_{\pmb{\slashed{k}}+\pmb{Q}, \alpha \beta} +  \sum_{\pmb{\slashed{k}} \notin FBZ} \epsilon^{E}_{\alpha \beta} (\pmb{\slashed{k}}) C^{\dagger}_{\pmb{\slashed{k}}, \alpha \beta}C_{\pmb{\slashed{k}}+\pmb{Q}, \alpha \beta} \\
= & \sum_{\pmb{k} \in FBZ} \epsilon^{E}_{\alpha \beta} (\pmb{k}) C^{\dagger}_{\pmb{k}, \alpha \beta}C_{\pmb{k}+\pmb{Q}, \alpha \beta} +  \sum_{\pmb{k} \in FBZ} \epsilon^{E}_{\alpha \beta} (\pmb{k+\pmb{Q}}) C^{\dagger}_{\pmb{k}+ \pmb{Q}, \alpha \beta}C_{\pmb{k}+2\pmb{Q}, \alpha \beta}
\end{align}
\end{widetext}

\noindent which after recognizing that $C_{\pmb{k}+2\pmb{Q}}= C_{\pmb{k}}$ gives the form in Eq. \ref{Eq:Conventioanal_1}.

\end{document}